\newcommand{\be}{\begin{equation}}
\newcommand{\bea}{\begin{eqnarray}}
\newcommand{\ee}{\end{equation}}
\newcommand{\eea}{\end{eqnarray}}
\newcommand{\nn}{\nonumber}

\newcommand{\qa}{\alpha}
\newcommand{\qb}{\beta}

\newcommand{\qG}{\Gamma}
\newcommand{\qd}{\delta}
\newcommand{\qD}{\Delta}
\newcommand{\qe}{\varepsilon}

\newcommand{\qh}{\eta}

\newcommand{\qy}{\theta}

\newcommand{\qk}{\kappa}
\newcommand{\ql}{\lambda}

\newcommand{\qr}{\rho}
\newcommand{\qs}{\sigma}

\newcommand{\qt}{\tau}

\newcommand{\qF}{\Phi}

\newcommand{\qJ}{\Psi}
\newcommand{\qo}{\omega}
\newcommand{\qO}{\Omega}

\newcommand{\tr}{{\rm tr}\,}

\newcommand{\tri}{\triangle}

\newcommand{\dagg}{^{\dag}}

\newcommand{\fr}[2]{{\textstyle \frac{#1}{#2}}}

\newcommand{\EE}{\mathop{{\mathbb{E}}}}

\newcommand{\RR}{{\mathbb R}}

\newcommand{\one}{{\mathbb 1}}
\newcommand{\sH}{{\sf H}}
\newcommand{\bits}{ \{0,1\} }
\newcommand{\Hmin}{{\sf H}_{\rm min}}
\newcommand{\Hmax}{{\sf H}_{\rm max}}

\newcommand{\cF}{{\mathcal F}}

\newcommand{\cH}{{\mathcal H}}
\newcommand{\cI}{{\mathcal I}}

\newcommand{\cK}{{\mathcal K}}

\newcommand{\cM}{{\mathcal M}}

\newcommand{\cO}{{\mathcal O}}

\newcommand{\cR}{{\mathcal R}}
\newcommand{\cS}{{\mathcal S}}
\newcommand{\cT}{{\mathcal T}}
\newcommand{\cU}{{\mathcal U}}

\newcommand{\cX}{{\mathcal X}}

\newcommand{\pr}{{\rm Pr}}

\newcommand{\isdef}{\stackrel{\rm def}{=}}

\newcommand{\mdel}{m_{\scriptscriptstyle\tri}}

\newcommand{\ket}[1]{| #1 \rangle}
\newcommand{\bra}[1]{\langle #1 |}

\documentclass[10pt]{article}
\usepackage{graphics}
\usepackage{graphicx}
\usepackage{url}
\usepackage{amsmath}
\usepackage{amssymb}
\usepackage{bbold}
\usepackage{a4wide}
\usepackage{enumitem}

\newtheorem{theorem}{Theorem}[section]

\newtheorem{proposition}[theorem]{Proposition}
\newtheorem{lemma}[theorem]{Lemma}
\newtheorem{definition}[theorem]{Definition}

\newtheorem{example}[theorem]{Example}

\begin{document}

\setlength{\parindent}{0mm}

\title{Can't Touch This: \\ unconditional tamper evidence from short keys}

\author{Bart van der Vecht$^1$, Xavier Coiteux-Roy$^2$, Boris \v{S}kori\'{c}$^1$\\
{\small $^1$Eindhoven University of Technology\quad $^2$Universit\`{a} della Svizzera Italiana, Lugano}
}

\date{ }

\maketitle

\begin{abstract}
\noindent 
Storing data on an external server with information-theoretic security, while using a key shorter than the data itself, is impossible. 
As an alternative, we propose a scheme that achieves information-theoretically secure tamper evidence: 
The server is able to obtain information about the stored data, but not while staying undetected.
Moreover, the client only needs to remember a key whose length is much shorter than the data.

\noindent
We provide a security proof for our scheme, based on an entropic uncertainty relation,
similar to QKD proofs.
Our scheme works if Alice is able to (reversibly) randomise the message to almost-uniformity
with only a short key.
By constructing an explicit attack we show that short-key unconditional tamper evidence
cannot be achieved without this randomisability.
\end{abstract}

\section{Introduction}
\label{sec:intro}

\subsection{Delegated Storage}
\label{sec:introdelegated}

Quantum information processing is markedly different from classical information processing.
For instance, performing a measurement on an unknown quantum state typically destroys state information.
Furthermore, it is impossible to clone an unknown state by unitary evolution \cite{WoottersZurek}.
Such properties are very interesting for security applications,
since they provide a certain amount of built-in confidentiality, unclonability and tamper-evidence.
Quantum physics also features entanglement of subsystems, which allows for feats like
teleportation \cite{teleport,MK2004} that have no classical analogue. 
The laws of quantum physics have been exploited in various security schemes, such as
Quantum Key Distribution (QKD) \cite{BB84,Ekert91,GP2000}, 
quantum anti-counterfeiting \cite{BBBW1982},
quantum Oblivious Transfer \cite{DFSS2005,Schaffner2010},
authentication and encryption of quantum states \cite{BCGST,Boykin2003,AMTW2000}, 
unclonable encryption \cite{uncl},
quantum authentication of PUFs \cite{QRIJQI,QRexp}, 
and quantum-secured imaging \cite{Malik2012}, to name a few.
For a recent overview of quantum-cryptographic schemes we refer to
\cite{BroadbentSchaffner2016}. 

In this paper we look at the problem of {\em Delegated Storage}.
Alice needs to store a large amount of data securely, but she does not have 
enough storage capacity herself.
The typical solution is to encrypt the data and then store it on a remote (`cloud') server Eve.
Since Alice has to remember the encryption key, this key is necessarily smaller than the data 
(otherwise Alice could have just stored the data herself).
It is well known that information-theoretic security is possible only when the key is at 
least as large as the entropy of the data.
Hence it is obvious that in Delegated Storage the confidentiality of the data
cannot be guaranteed unconditionally, not even using quantum physics. 
A computationally unbounded Eve will always be able to extract information about the data from the (quantum) ciphertext.

We show that,
somewhat surprisingly, it {\em is} possible in Delegated Storage
to get information-theoretic guarantees for a security property other than confidentiality:
tamper evidence (tampering detection).
We present a quantum Delegated Storage scheme for classical data
which makes it impossible for Eve to learn anything about Alice's data without alerting Alice,
even if Eve has unbounded powers of (quantum) computation, measurement, storage etc.
Our scheme is close in spirit to QKD, and in fact it is useful to imagine 
Delegated Storage as a sort of QKD where Bob is `future Alice' who retrieves and decrypts the stored cipherstate,
and storage on the server corresponds to travelling qubits.
There are some subtle differences with QKD, however, namely (i) the short encryption key,
(ii) the availability of the ciphertext at the moment when Eve attacks the qubits,
and 
(iii) Bob's inability to send any message to Alice.
These subtle differences conspire to necessitate a security proof that differs nontrivially from QKD security proofs, 
though many
well known ingredients can be re-used.

\subsection{Related work}
\label{sec:relatedwork}

Several works have appeared on the topic of provable deletion of remotely stored data.
Coiteux-Roy and Wolf \cite{CW2019} introduced the task of 
Delegated Storage and provable deletion with a short-key requirement for both tasks.
However, they did not settle the question whether unconditional tamper evidence is achievable.
Independently, Broadbent and Islam \cite{BI2020} achieved information-theoretic security for provable deletion
using keys that are as long as the message.

L\"{u}tkenhaus, Marwah and Touchette~\cite{LMT2020} use a form of Delegated Storage to store a fully-randomised bit commitment
on temporarily trusted servers, with the possibility of recall.
They don't require a short key in their definition and use a key as long as the message in their protocol.

The verification process in Delegated Storage involves the measurement of a quantum state by the verifier;
the prover has to send this quantum state to the verifier.
This is different from Provable Deletion protocols and from 
Molina, Vidick and Watrous's tickets variant \cite{MVW2012} of Wiesner's quantum money,
where the stored data is quantum but 
the communication between the prover and the verifier is classical during the verification phase.

\subsection{Contributions and outline}
\label{sec:contrib}

\begin{itemize}[leftmargin=4mm,itemsep=0mm,topsep=1mm]
\item
We define {\em Correctness}, {\em Security} and {\em Usefulness} for Delegated Storage.
Correctness means that, in case of low disturbance of the stored quantum states, Alice should not get alerted and should 
be able to recover the message.
Security means that Eve cannot learn a non-negligible amount of information about the stored message without
alerting Alice.
(This definition {\em does} allow Eve to learn the full message while alarming Alice.)
Usefulness means that Alice's locally stored data is smaller than the remotely stored message.
\item
We present {\sc Can'tTouchThis},
our Delegated Storage scheme.
As a first step Alice derives, in a reversible way, an almost-uniform string $m$ from the message~$\mu$.
Our scheme requires that this randomisation step is possible without the introduction of long keys;  
hence the entropy of~$\mu$ must be sufficiently high to allow for using an extractor,
or Alice must know the distribution of~$\mu$ with sufficient accuracy in order to apply compression-based
randomisation techniques.
Then,
Alice extracts a one-time pad from a random string $x$; the $x$ is encoded into qubits.
She computes a ciphertext by masking $m$ with the one-time pad.
She stores the ciphertext and the qubits on the server.
In between the qubits that contain $x$ there are `trap' qubits in random positions.
When Alice recovers the stored data, she inspects these trap states to see if they have changed.
\item
We prove that our scheme satisfies the Correctness and Security properties.
If $\ell$ is the message length and $\qb_0$ the bit error rate of the quantum channel, 
then asymptotically $n= \frac{\ell}{1-h(\qb_0)}$ qubits are required\footnote{
$h$ is the binary entropy function.
}, and 
Alice has to remember a syndrome of (asymptotic) size $\ell \frac{h(\qb_0)}{1-h(\qb_0)}$;
the syndrome is the main `key' that she has to store locally.
{\sc Can'tTouchThis} allows
the message to be longer than the key only when $1-2h(\qb_0)>0$. 
This inequality is familiar in Quantum Key Distribution, where it represents the condition
for having positive key rate without two-way communication. 
\item
We propose a method for recursively applying {\sc Can'tTouchThis}. 
The syndrome is not stored locally, but using {\sc Can'tTouchThis}.
The effect is that Alice has to remember a shorter key; asymptotically the number of qubits 
stored on the server is $n\to\frac{\ell}{1-2h(\qb_0)}$.
This expression too is familiar from QKD, where it stands for the number of qubits required to
generate a key of length $\ell$.
\item
Our scheme needs a preprocessing step to reversibly transform the message $\mu$ into an almost-uniform string $m$
which then serves as the `message' in the quantum part of the protocol. 
We show that this need for a uniform input is not a deficiency of our scheme or our proof technique, but in fact a
fundamental requirement.
We introduce an attack called {\sc Support} which tries to determine one bit: 
whether the plaintext is the one with the highest a-priori probability.
We consider delegated storage in general {\em without preprocessing}
and lowerbound the advantage that {\sc Support} yields as a function of the key length and
the min-entropy of the plaintext.
This lower bound serves as a kind of `no go' theorem:
In the case of a low min-entropy distribution that is not known to Alice,
our bound implies that the Security property cannot be achieved with a short key.
\item
We propose two ways in which to achieve a reduced form of tamper evidence in case of the `no go'
situation mentioned above.
(i) Introducing a temporary computational assumption; 
(ii) secret sharing over multiple servers, with the temporary assumption that they are not all colluding.
\end{itemize}
The outline is as follows.
In Section~\ref{sec:prelim} we introduce notation and list useful definitions and lemmas.
The security definition is given in Section~\ref{sec:securitydef}.
In Section~\ref{sec:protocol} we describe {\sc Can'tTouchThis}, and in Section~\ref{sec:proof}
we do the security analysis.
Section~\ref{sec:parameters} discusses parameter settings and the recursive scheme.
In Section~\ref{sec:imposs} we prove the `no go' result for low-minentropy distributions that are not known to Alice.
In Section~\ref{sec:discussion} we discuss alternative scheme constructions
and weaker schemes in the `no go' situation.

\section{Preliminaries}
\label{sec:prelim}

\subsection{Notation and terminology}
\label{sec:notation}

Sets are written in calligraphic font.
Classical Random Variables (RVs) are denoted with capital letters, and their realisations
with lowercase letters. 
The expectation with respect to $X$ is denoted as 
$\EE_x f(x)=\sum_{x\in\cX}\pr[X=x]f(x)$.
The notation $X\sim P$ means that $X$ has distribution~$P$. We then write $P(x)=\pr[X=x]$.
The statistical distance between two RVs $X,Y\in\cX$, with $X\sim P$ and $Y\sim Q$, is given by 
$\qD(X,Y)=\fr12\sum_{x\in\cX}|P(x)-Q(x)|$.

Bitwise XOR of binary strings is written as `$\oplus$'.
For the first $\ell$ bits of the string $s$ we write $s_{[1:\ell]}$. 
The Hamming weight of $s$ is denoted as $|s|$.
The notation `$\log$' stands for the logarithm with base~2.
The function $h$ is the binary entropy function $h(p)=p\log\fr1p+(1-p)\log\fr1{1-p}$.

The Kronecker delta is denoted as $\qd_{ab}$.
We will speak about
`the bit error rate $\qb$ of a quantum channel'.
This is defined as the probability that a classical bit $x$, sent by Alice embedded in a qubit,
arrives at Bob's side as the flipped value $\bar x$. 

For quantum states we use Dirac notation.
The notation `tr' stands for trace.
$H=\fr1{\sqrt2}{1 \; \phantom{-}1\choose  1 \; -1}$ is the Hadamard matrix.
Let $A$ be a matrix with eigenvalues~$\ql_i$. 
The 1-norm of $A$ is written as $\|A\|_1=\tr\sqrt{A\dagg A}=\sum_i|\ql_i|$. 
The trace norm is $\|A\|_{\rm tr}=\fr12 \|A\|_1$.
Quantum states with non-italic label `A', `B' and `E' indicate the subsystem of Alice/Bob/Eve.
The space of normalised mixed states on Hilbert space $\cH$ is written as $\cS(\cH)$.
More generally, the space of sub-normalised states is $\cS_\leq(\cH)$.

Consider classical variables $X,Y$ and a quantum system under Eve's control that depends on $X$ and~$Y$. 
The combined classical-quantum state is $\qr^{XY \rm E}=\EE_{xy} \ket{xy}\bra{xy} \otimes \qr^{\rm E}_{xy}$. 
The state of a sub-system is obtained by tracing out all the other subspaces, 
e.g. $\qr^{ Y \rm E}={\rm tr}_X \qr^{XY\rm E}=\EE_y \ket y\bra y\otimes\qr^{\rm E}_y$, with $\qr^{\rm E}_y=\EE_{x|y}\qr^{\rm E}_{xy}$.
The fully mixed state on Hilbert space $\cH_A$ is denoted as~$\chi^A$.
A general measurement is described as a Positive Operator Valued Measure (POVM). 
A POVM $\cM$ on a Hilbert space~$\cH$,
with a set of possible measurement outcomes $\cU$,
consists of positive semidefinite operators $\{M_u\}_{u\in\cU}$ acting on $\cH$ that satisfy
$\sum_{u\in\cU}M_u=\one$.
The probability of outcome $u$ is $\tr M_u \qr$.

We define the rate of a quantum communication protocol as the number of message bits communicated per sent qubit.

\subsection{Definitions and lemmas}
\label{sec:prelimlemmas}

Here we present a number of definitions and lemmas that will be used in the remainder of the paper.

\begin{definition}[R\'{e}nyi entropy]
\label{def:Renyi}
Let $\cX$ be a discrete set.
Let $X\in\cX$ be a classical variable.
Let $\qa\in(0,1)\cup(1,\infty)$.
The R\'{e}nyi entropy of order $\qa$ is denoted as $\sH_\qa(X)$
and is defined as
\be
	\sH_\qa(X)=\frac{-1}{\qa-1}\log\sum_{x\in\cX}(\pr[X=x])^\qa.
\ee
\end{definition}

\begin{definition}[Smooth R\'{e}nyi entropy]
\label{def:smoothRenyi}
Let $X\sim P$ be a discrete classical variable.
Let $\qa\in(0,1)\cup(1,\infty)$.
Let $\qe\geq 0$.
The $\qe$-smooth R\'{e}nyi entropy of order $\qa$ is denoted as $\sH_\qa^\qe(X)$
and is defined as
\be
	\sH_\qa^\qe(X) = \max_{Y\sim Q,\; Q\in B^\qe(P)} \sH_\qa(Y), 
\ee
where $B^\qe(P)$ is a sub-normalised vicinity of $P$ such that
for $Q\in B^\qe(P)$ it holds that 
$\sum_{x\in\cX}Q(x)\geq 1-\qe$
and
$\forall_{x\in\cX}\;Q(x)\leq P(x)$.
\end{definition}

\begin{definition}[Min-entropy]
Let $\qr^{AB}\in\cS(\cH_A\otimes\cH_B)$ be a bipartite state.
The min-entropy of the subsystem A conditioned on B is denoted as $\Hmin(A|B)_\qr$ and defined as 
\be
	\Hmin(A|B)_\qr = \max_{\qs\in\cS(\cH_B)}\sup\{\ql\in\RR: 2^{-\ql}\one_A\otimes\qs\geq\qr^{AB}  \}.
\ee
\end{definition}

When the subsystem $A$ is classical, the min-entropy is related to the guessing entropy.

\begin{definition}[Guessing entropy given quantum side information]
Let $\qr^{X\rm E}$ be a state where $X\in\cX$ is classical.
The guessing probability of $X$ given the subsystem E is denoted as $p_{\rm guess}(X|{\rm E})_\qr$
and defined as
\be
	p_{\rm guess}(X|{\rm E})_\qr = \max_\cM \sum_{x\in\cX}\pr[X=x] \tr(M_x \qr^{\rm E}_x),
\ee
where $\cM$ is a POVM given by operators $(M_x)_{x\in\cX}$.
\end{definition}

\begin{definition}[Min-entropy of a classical variable given quantum side information]
Let $\qr^{X\rm E}$ be a state where $X$ is classical.
The min-entropy of $X$ given the subsystem E is denoted as $\Hmin(X|{\rm E})_\qr$
and is defined as
\be
	\Hmin(X|{\rm E})_\qr = -\log p_{\rm guess}(X|{\rm E})_\qr.
\ee
\end{definition}

\begin{definition}[Smooth min-entropy]
\label{def:smoothHmin}
(See e.g.\,\cite{TCR2010})
Let $\qr^{AB}$ be a bipartite state.
The $\qe$-smooth min-entropy of A given B is denoted as $\Hmin^\qe(A|B)_\qr$ and is defined as\footnote{
For the distance metric between $\qr$ and $\qt$, \cite{TCR2010} actually uses 
the {\em generalised trace distance} or
{\em purified distance} 
$\|\qt-\qr\|_{\rm p}=\fr12\|\qt-\qr\|_1+\fr12 |\tr\qt-\tr\qr|$.
This distinction is not important in the current paper.
}
\be
	\Hmin^\qe(A|B)_\qr = \max_{\qt:\; \| \qt-\qr\|_1\leq\qe} \Hmin(A|B)_\qt.
\ee
\end{definition}

\begin{definition}[Smooth max-entropy]
\label{def:smoothHmax}
(See e.g.\,\cite{TCR2010})
Let $\qr^{AB}$ be a state, and let $\qr^{ABC}$ be a purification of $\qr^{AB}$.
The smooth max-entropy of $A$ given the subsystem B is denoted as $\Hmax^\qe(A| \rm B)_\qr$ and is defined as
\be
	\Hmax^\qe(A| \rm B)_\qr = -\Hmin^\qe(A|C)_\qr.
\ee
\end{definition}
The max-entropy of a classical variable $X$ is a measure of the size of the support of~$X$.

\begin{definition}[Extractor]
Let $f: \bits^n\times\bits^d\to\bits^\ell$ be a function.
Let $R\in\bits^d$ be a uniformly random seed.
Let $U\in\bits^\ell$ be a uniform RV.
The function $f$ is called a $(k,\qe)$-extractor if 
\be
	\Hmin(X)\geq k \implies \qD\big(f(X,R),U\big)\leq \qe.
\ee
\end{definition}

\begin{definition}[Strong extractor]
Let $f: \bits^n\times\bits^d\to\bits^\ell$ be a function.
Let $R\in\bits^d$ be a uniformly random seed.
Let $U\in\bits^\ell$ be a uniform RV.
The function $f$ is called a $(k,\qe)$ strong extractor if 
\be
	\Hmin(X)\geq k \implies \qD\big(R f(X,R),R U\big)\leq \qe.
\ee

\end{definition}

\begin{definition}[Quantum-proof strong extractor]
\label{def:qproofExt}
(See e.g.\,\cite{DPVR2012}.)
Let $f: \bits^n\times\bits^d\to\bits^\ell$ be a function.
Let $R\!\in\!\bits^d$ be a uniformly random seed.
Let $X\!\in\!\bits^n$ be a classical RV and let $\qr^{X\rm E}$ be a classical-quantum system
comprising the classical $X$ entangled with a quantum system `E'. 
Let $Z=f(X,R)$.
The function $f$ is called a quantum-proof $(k,\qe)$ strong extractor if 
\be
	\Hmin(X|E)_\qr\geq k 
	\quad\implies \quad
	\fr12 \big\| \qr^{ZR\rm E}-\chi^Z\otimes\chi^R\otimes \qr^{\rm E} \big\|_1\leq \qe.
\ee
\end{definition}

\begin{definition}[Universal hash]
A family of hash functions $\cF=\{f:\cX\to\cT  \}$ is called two-universal (or universal)
if for all distinct pairs $x,x'\in\cX$ it holds that $\pr_{f\in\cF}[f(x)=f(x')]=1/|\cT|$.
Here the probability is over random~$f\in \cF$.
\end{definition}

A universal hash function is a strong extractor.

\begin{lemma}[Leftover Hash Lemma] (See \cite{VTOSS10}.) 
\label{lemma:LHL}
Let $X\in\cX$ be a random variable.
Let $f:\cR\times \cX\to\bits^\ell$ be a universal hash function.
Let $U\in\bits^\ell$ be a uniform variable.
Then 
\be
	\ell\leq \max_{\qh\in[0,\qe)}\Big[\sH^\qh_2(X)+2-\log\frac1{\qe(\qe-\qh)}\Big] 
	\quad\implies\quad \qD(Rf(R,X),RU)\leq\qe.
\ee

\end{lemma}

\begin{lemma}
\label{lemma:invertibleHash}
Let $F:\bits^\nu\times\bits^\nu\to \bits^\nu$ be given by $F(w,x)=w\cdot x$,
where the multiplication is in GF$(2^\nu)$.
Let $\ell\leq\nu$. 
Let $\qF:\bits^\nu\times\bits^\nu\to\bits^\ell$ be constructed as 
$\qF(w,x)\isdef F(w,x)[1:\ell]$, i.e. the first $\ell$ bits of $w\cdot x$.
Then $\qF$ is a universal hash.
\end{lemma}
\underline{\it Proof:}
Let $x,x'\in GF(2^\nu)$, with $x'\neq x$. A collision $\qF(x')=\qF(x)$
occurs when $((x+x')w)[1:\ell]=0$.
The number of values $w$ for which this can occur is $2^{\nu-\ell}$;
the probability of drawing such $w$ is $2^{\nu-\ell}/2^\nu=1/2^\ell$.
\hfill$\square$

\begin{lemma}
\label{lemma:existExt}
{\bf (Existence of quantum-proof strong extractor.)}
(Based on Theorem~9 in \cite{TSSR2011}.)
There exists a quantum-proof 
$(\ell+4\log\fr1\qe-2,\qe)$-strong extractor 
from $\bits^n$ to $\bits^\ell$
with uniform seed of length~$n$.
\end{lemma}


\begin{lemma}[Entropic uncertainty relation for BB84 bases.]
\label{lemma:entropicBB84}
(See \cite{TR2011})
Let $\qr^{\rm ABE}$ be any state of the three-partite system ABE,
where the subsystem A consists of $n$ qubits.
Let $X\in\bits^n$ be the outcome of a measurement on A in the standard basis.
Let $X'\in\bits^n$ be the outcome of a measurement on A in the Hadamard basis.
Then
\be
	\Hmin^\qe(X|{\rm E})_\qr +\Hmax^\qe(X'|{\rm B})_\qr \geq n.
\ee
\end{lemma}


\begin{lemma}
\label{lemma:sampling}
(Lemma~6 in~\cite{TL2017}.)
Let $z\in\bits^{n+r}$. Let $\qb,\nu>0$.
Let $\cI\subset[n+r]$, with $|\cI|=r$, be a uniformly distributed random variable representing a choice of $r$ out of
$n+r$ positions.  
Then
\be
	\pr\Big[\sum_{i\in\cI}z_i\leq r\qb \;\;\wedge\;\; \sum_{i\notin\cI}z_i\geq n(\qb+\nu)\Big]
	\leq e^{-2\nu^2 r\frac{nr}{(n+r)(r+1)}}.
\ee
\end{lemma}

\begin{lemma}
\label{lemma:Hoeffding}
{\bf(Hoeffding inequality.)}
Let $X_1,\ldots,X_n\in\bits$ be $n$ i.i.d. Bernoulli variables with parameter $p$.
Then
\be
	\pr\Big[\sum_{i=1}^n X_i \geq n(p+\qe)\Big] \leq e^{-2\qe^2 n}.
\ee
\end{lemma}

\begin{definition}[Message Authentication Code.]
\label{def:MAC}
A Message Authentication Code (MAC) 
with message space $\cM$ and key space $\cK$
consists of a pair
(Tag,Ver), where 
Tag is a probabilistic poly-time algorithm that, given a key $k\in\cK$ and a message $m\in\cM$, returns a tag $\qy\leftarrow {\rm Tag}(k,m)$;
Ver is a deterministic poly-time algorithm that, on input $k,m,\qy$, outputs 
${\rm Ver}(k,m,\qy)\in$ \{`accept',`reject'\}.
It must hold that 
$\forall_{m\in\cM,k\in\cK}\; 
{\rm Ver}(k,m,{\rm Tag}(k,m))=\mbox{`accept'}$.
\end{definition}

When the MAC key $k$ is used only once,
it is possible to achieve information-theoretic security \cite{WegmanCarter1981,Stinson}.
A one-time MAC is said to be $\qe$-secure if an adversary, upon observing a pair $(m,\qy)$,
cannot create a valid forgery $(m',\qy')$, with $m'\neq m$, except with probability~$\qe$.
Constructions are known with short tags and short keys.
For instance, a construction by den Boer \cite{denBoer1993} achieves $\qe$-security with
key size $\approx 2\log\fr1\qe+2\log\log|\cM|$ and tag size $\approx \log\fr1\qe+\log\log|\cM|$.

\section{Attacker model and security definition}
\label{sec:securitydef}

We adopt the attacker model that is customary in QKD.
No information leaks from Alice's lab, i.e. there are no side channels. 
Eve has unlimited (quantum) computational resources and is able to perform any measurement allowed by theory. 
All noise on the quantum channel is considered to be caused by Eve.

Alice draws her message from a certain probability distribution.
We will consider three scenarios,
\begin{enumerate}[leftmargin=5mm,itemsep=0mm,topsep=1mm]
\item
{\bf Fully randomised}. 
From Eve's point of view, the message is uniformly distributed.
\item
{\bf Weakly randomised}.
From Eve's point of view, the message is not uniformly distributed.
However, the distribution has certain favourable properties, and
Alice has sufficient knowledge of it to construct an almost-uniform string 
from the message, by using e.g.\,an extractor (Lemma~\ref{lemma:LHL}), 
prefix coding techniques (Section~\ref{sec:design}) or a combination.
\item
{\bf Non-randomised}.
None of the above apply.
\end{enumerate}
We will show that Delegated Storage can be achieved in the 1st and 2nd scenario, while
for a special case of the 3rd scenario we will prove a `no-go' theorem. 

Security proofs are often given for the EPR-based version of a protocol.
We will follow the same approach. 
In this section we define, in the EPR setting, what is meant by `security' for a Delegated Storage protocol.

\vskip2mm

\underline{The semantics}.\\
The classical variables in the protocol can be abstractly grouped as:
The message $M$, the set of keys $K$, the data $R$ that Alice has to remember apart from the keys,
the transcript $T$ (classical data stored on the server), the modified transcript $T'$ retrieved by Alice,
a binary flag $\qO\in\bits$ indicating {\tt accept} ($1$) or {\tt reject} ($0$),
and the reconstructed message $\hat M$.
The input to the protocol consists of EPR states and the classical $M$,$K$.
The final output is a quantum-classical state 
$\qr^{M\hat M T\qO \rm E}$
containing the classical subsystems
$M$, $\hat M$, $T$, $T'$, $\qO$ and Eve's quantum side information. We denote Eve's system as `E'.
The output state can be written as
$\qr^{M\hat M TT'\qO \rm E}= \qr^{M\hat M TT' \rm E}_{[\qo=0]}+\qr^{M\hat M TT' \rm E}_{[\qo=1]}$,
with $\tr \qr^{M\hat M TT' \rm E}_{[\qo=0]}=\pr[\qO=0]$
and $\tr \qr^{M\hat M TT' \rm E}_{[\qo=1]}=\pr[\qO=1]$.
Furthermore we write
$\qr^{M\hat M TT' \rm E}_{[\qo=0]}=\pr[\qO=0]_\qr \qr^{M\hat M TT' {\rm E}|\qO=0}$
and similarly for $\qo=1$.

\vskip2mm
\underline{Correctness}.\\
We say that the Delegated Storage protocol is $\qe$-correct if the following holds,\\
\be
	\mbox{Eve behaves honestly }\quad\implies\quad
	\pr[\qO=1 \;\wedge\; \hat M=M] \geq 1-\qe.
\label{defcorrect}
\ee

\vskip2mm
\underline{Security}.\\
We say that the Delegated Storage protocol is $\qe$-secure if the following statement holds.
\be
	\Big\| \qr^{MTT'\rm E}_{[\qo=1]}-\EE_m\ket m\bra m\otimes\qr^{TT'\rm E}_{[\qo=1]}  \Big\|_1\leq\qe.
\label{secstatement}
\ee
Eq.(\ref{secstatement}) can be read as:
``If $\pr[\qO=1]$ is negligible then we are making no demands.
If $\pr[\qO=1]$ is non-negligible then we demand that $M$ is decoupled from Eve''.
Note that security properties formulated in terms of the 1-norm (or trace norm) are
{\em composable} \cite{Can2001,ben2005universal,RK2005,fehr2009composing}
with other (sub-)protocols.

\vskip2mm
\underline{Usefulness}.\\
Let the message space be $\cM$ and the key space $\cK$.
We define the {\em usefulness} parameter $Y\leq 1$ as 
\be
	Y=\frac{|\cM|-|\cK|}{|\cM|}.
\ee
A positive usefulness means that the amount of data that Alice has to store locally is
less than the message size.

\section{Our Delegated Storage protocol {\sc Can'tTouchThis}}
\label{sec:protocol}

\subsection{Design considerations: message randomisation}
\label{sec:design}

The potential messages $\mu$ that Alice may store come from a message space $\cM$.
The probability distribution may be far from uniform on~$\cM$.
We will see later on that our scheme requires the stored message to be close to uniform,
and that near-uniformity is in fact a necessary condition for Delegated Storage in general.
For non-uniform messages one runs into the problem that the ciphertext, which is visible to Eve when she attacks the quantum state,
causes leakage about the data contained in the qubits.

Alice hence needs to transform $\mu$ into an almost-uniform string $M$
while remembering only a limited amount of information for recovery purposes.
She does this in two steps.
First she applies a {\em prefix code}\footnote{
In a prefix code, messages from $\cM$ get compressed to different sizes depending on their probability; likely messages
are assigned short representations. 
Furthermore, no  codeword exists that is the prefix of another codeword.
When parsing from the left, the end of a codeword is always recognized without any need for 
an `end of string' delimiter.
}
(see e.g. \cite{Huffman1952,Elias1975}) to losslessly compress~$\mu$.
The codeword is padded with random bits so that every $\mu\in\cM$ that has nonzero probability of occurring
is transformed into a string $M_0$ of fixed length~$\ell_0$.
The $\ell_0$ depends only on the probability distribution;
it is the length of the longest codeword\footnote{
In some extreme cases, such as Example~\ref{ex:hmin}, it may happen that $\ell_0$ is slightly larger than $\log|\cM|$,
in which case it is not really a compression.
}
in the prefix code.
The fact that the code is a {\em prefix code} ensures that the start of the padding can be
recognized. 
Hence, the randomisation comes `for free': Alice does not have to remember the padding bits.

The second step is to apply an invertible strong extractor to~$M_0$.
(See Lemma~\ref{lemma:invertibleHash}).
This maps $M_0\in\bits^{\ell_0}$ to $M\in\bits^\ell$.
If one is willing to tolerate non-uniformity $\qe_0$ then, according to Lemma~\ref{lemma:LHL},
the extractable randomness is
\be
	\ell\isdef \max_{\qh\in[0,\qe_0)}\Big[\sH^\qh_2(M_0)+2-\log\frac1{\qe_0(\qe_0-\qh)}\Big].
\label{extractable}
\ee
Alice needs to store locally $\ell_0-\ell$ secret bits in order to later recover $M_0$ from~$M$.
Once all this is in place, Alice applies a Delegated Storage method 
that can straightforwardly be proven secure when the message is uniform.

Note that the prefix code method makes sense only if Alice has a reasonably precise knowledge of the
distribution $P$ of~$\mu$.
The step with the strong extractor requires less knowledge: only a correct estimate of
$\sH_2(M_0)$ is required. 
Such an estimate is feasible e.g.\,when $P$ is drawn (in a potentially unknown way)
from an ensemble of distributions which each have a known lower bound on the $\sH_2$ entropy.

An example of prefix coding plus padding is shown below, for a rather extreme distribution.
Here the result $M_0$ is practically uniform, but this will not be the case in general.
Example~\ref{ex:hmin} shows that it is sometimes possible to randomise the input very effectively `for free'
even when the $\sH_2$-entropy is very low; the randomisation succeeds because Alice knows the distribution of $\mu$.

\begin{example}
\label{ex:hmin}
Consider the following probability distribution on $\cM=\bits^L$.
For one string $\mu_0$ it holds that $\pr[\mu=\mu_0]=\fr12$; all other strings have
probability $\frac{1/2}{2^L-1}$.
This distribution has $\Hmin(\mu)=1$, collision entropy $\sH_2(\mu)=2+\cO(2^{-L})$
and Shannon entropy $\sH(\mu)=\frac L2+\cO(1)$. 
The prefix code with padding is constructed as follows.
A string $\mu\neq\mu_0$ is encoded as $(0||\mu)\in\bits^{L+1}$.
The string $\mu_0$ is encoded as `1' followed by $L$ random padding bits.
The resulting string $M_0\in\bits^{L+1}$ has the probability distribution
\be
	\pr[M_0 = (0||x) ] = (1-\qd_{x,\mu_0}) \frac{1/2}{2^L-1};
	\quad\quad\quad
	\pr[M_0 = (1||x)] = 2^{-(L+1)}
\ee
for any $x\in\bits^L$.
It has min-entropy $\Hmin(M_0)=\log(2^L-1)+1>L+1-2^{-L}/\ln2$.
\end{example}

\subsection{Protocol steps}
\label{sec:protocolsteps}

\underline{Setup phase}.\\
Alice chooses $\qe_0$ and sets $\ell$ according to (\ref{extractable}).
She chooses parameter values  $r$ (number of trap positions), 
and $\qe_{\rm qp}$
(security of the quantum-proof extractor). 

We consider a noise model where (in the case of passive Eve) each qubit independently suffers from noise in the same way.
Let $\qb_0$ be the quantum bit error rate.\footnote{
i.e. if Alice prepares a qubit state in a certain basis and later retrieves the qubit from storage and measures it in the same basis,
there is a probability $\qb_0$ that the bit value has flipped.
}
Alice sets a parameter $\qb$, with $\qb$ slightly larger than~$\qb_0$. (In the limit of large message length, $\qb$ will go to~$\qb_0$.)
Furthermore, Alice chooses an Error-Correcting Code $C$ that is able to deal with error rate~$\qb+\nu$.
Here $\nu$ is a small positive parameter. (Asymptotically $\nu$ goes to zero.) 
The ECC message length is denoted as $\qk$ and the codeword length as~$n$.
We write the syndrome function as ${\tt Syn}: \bits^n\to\bits^{n-\qk}$, and the syndrome decoding as 
${\tt SynDec}: \bits^{n-\qk}\to\bits^n$.
The ECC message length is set as
\be
	\qk =  \ell+4\log\frac1{\qe_{\rm qp}}-2.
\label{setECC}
\ee
Alice chooses a quantum-proof $(\qk,\qe_{\rm qp})$-strong extractor $f:\bits^d\times\bits^n\to\bits^{\ell}$,
where $d$ is the length of the seed.

 Alice has an invertible randomisation function ${\tt Compress}:\cM\to\bits^{\ell_0}$.
This includes the random padding step.
The corresponding inverse function is ${\tt Decompress}:\bits^{\ell_0}\to\cM$,
which includes discarding the padding bits.

Alice chooses an $\qe_{\rm mac}$-secure one-time Message Authentication Code
with tag function $\qG: \cK\times \bits^{\ell_0+d+\ell}\to \bits^{\ql}$.

\vskip2mm

\underline{Message preparation}.\\
Alice has a message $\mu\in\cM$. She performs the following steps.
\begin{enumerate}[leftmargin=5mm,itemsep=0mm,topsep=1mm]
\item 
$m_0={\tt Compress}(\mu)$.
\item
Draw random 
seed $w\in\bits^{\ell_0}$.
Compute $p=w\cdot m_0$, where the multiplication `$\cdot$' is in GF$(2^{\ell_0})$. 
Parse $p$ as $p=m\| \mdel$, with $m\in\bits^{\ell}$.
\end{enumerate}

\vskip2mm

\underline{Encryption and storage}.
\begin{enumerate}[leftmargin=5mm,itemsep=0mm,topsep=1mm]
\setcounter{enumi}{2}
\item 
Draw random strings $\xi,t\in\bits^{n+r}$  with $|t|=r$. 
(The string $t$ defines a subset $\cT\subset[n+r]$ of size $r$ which points at the trap locations.)
Create strings $v=\xi_\cT$ (trap values) and $x=\xi_{[n+r]\setminus\cT}$ (payload). 
Prepare the quantum state $\ket\qJ=\bigotimes_{j=1}^{n+r} H^{t_j}\ket{\xi_j}$.
\item
Draw random seed $u\in\bits^d$. 
Compute syndrome $s={\tt Syn}\,x$, one-time-pad $z=f(u,x)$ and ciphertext
$c=m\oplus z$.
\item
Draw random MAC key $\qh\in\cK$. 
Compute the tag $\qy=\qG(\qh,w\| u\| c)$.
Store $w,u,c,\qy$ and $\ket\qJ$ on the server. 
Remember $\qh,\cT,v,s,\mdel$.
Forget all other variables.
\end{enumerate}

\vskip2mm

\underline{Testing and decryption}.
\begin{enumerate}[leftmargin=5mm,itemsep=0mm,topsep=1mm]
\setcounter{enumi}{5}
\item 
Retrieve classical data $w',u',c',\qy'$ and state $\ket{\qJ'}$.
Set $\qo=0$.
If $\qy' \neq \qG(\qh,w'\|u'\| c')$ then abort.
\item
In positions $\cT$ measure $\ket{\qJ'}$ in the Hadamard basis; in all other positions in the standard basis.
The measurement result is $v'\in\bits^r$ from the trap locations and $x'\in\bits^n$ from the other locations.
If $|v'\oplus v|> \qb r$ then abort.
\item
Reconstruct $\hat x=x'\oplus{\tt SynDec}(s\oplus{\tt Syn}\, x')$. 
If {\tt SynDec} fails then abort, else continue.
Set $\qo=1$.
Compute $\hat z=f(u',\hat x)$ and $\hat m=\hat z\oplus c'$.
\item
Compute $\hat m_0=(w')^{-1}\cdot(\hat m\|\mdel)$ and 
$\hat\mu={\tt Decompress}(\hat m_0)$.
\end{enumerate}


\begin{table}[h]
\begin{center}
{\small
\begin{tabular}{|l|l|}
\hline
{\it Notation} & {\it Meaning}
\\ \hline
$\qb_0,\qb$ & actual and accepted bit error rate on the quantum channel
\\ \hline
$c$ & classical ciphertext
\\ \hline
$d$ & seed length
\\ \hline
$\qe_0$ & uniformity parameter of the strong extractor
\\ \hline
$\qe_{\rm mac}$ & security of the MAC
\\ \hline
$\qe_{\rm qp}$ & uniformity parameter of the quantum-proof strong extractor
\\ \hline
$\qh$ & MAC key
\\ \hline
$f$ & quantum-proof strong extractor
\\ \hline
$\qG$ & MAC function
\\ \hline
$H$ & Hadamard operator
\\ \hline
$\qk$ & ECC message length
\\ \hline
$\ell$ & output length of quantum-proof strong extractor
\\ \hline
$\ell_0$ & lossless compression length
\\ \hline
$\ql$ & length of authentication tag
\\ \hline
$m_0$ & compressed message
\\ \hline
$m$ & output of strong extractor
\\ \hline
$\mdel$ & stored for strong extractor inversion
\\ \hline
$\mu$ & original message
\\ \hline
$n$ & number of qubits that carry a payload
\\ \hline
$\nu$ & small positive parameter
\\ \hline
$\qo$ & success flag
\\ \hline
$r$ & number of trap qubits
\\ \hline
$s$ & syndrome of payload $x$
\\ \hline
{\tt Syn} & syndrome function
\\ \hline
$t$ & string indicating the trap positions
\\ \hline
$\cT$ & set of trap positions
\\ \hline
$\qy$ & authentication tag
\\ \hline
$u$ & seed for the quantum-proof extractor
\\ \hline
$v$ & trap values
\\ \hline
$w$ & seed for the strong extractor
\\ \hline
$x$ & payload in the non-trap positions
\\ \hline
$z$ & classical one-time pad
\\ \hline
\end{tabular}
}
\caption{\it Notation.}
\end{center}
\end{table}

\clearpage

\begin{figure}[h]
\begin{center}
\includegraphics[width=0.45\textwidth]{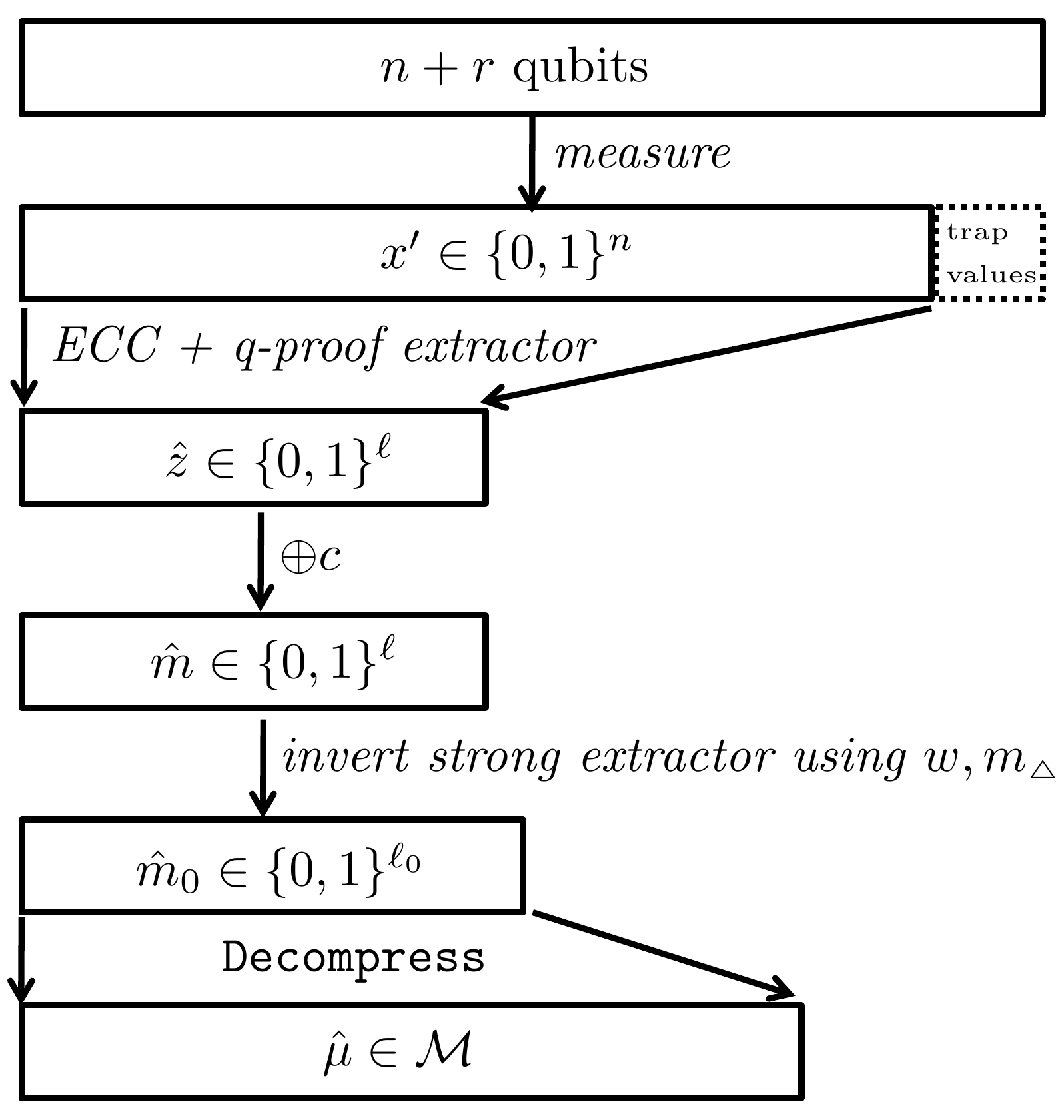}
\caption{\it Visualisation of the variables and the steps in the reconstruction of the message~$\mu$.}
\label{fig:data}
\end{center}
\end{figure}

\begin{figure}[h]
\begin{center}
\includegraphics[width=0.8\textwidth]{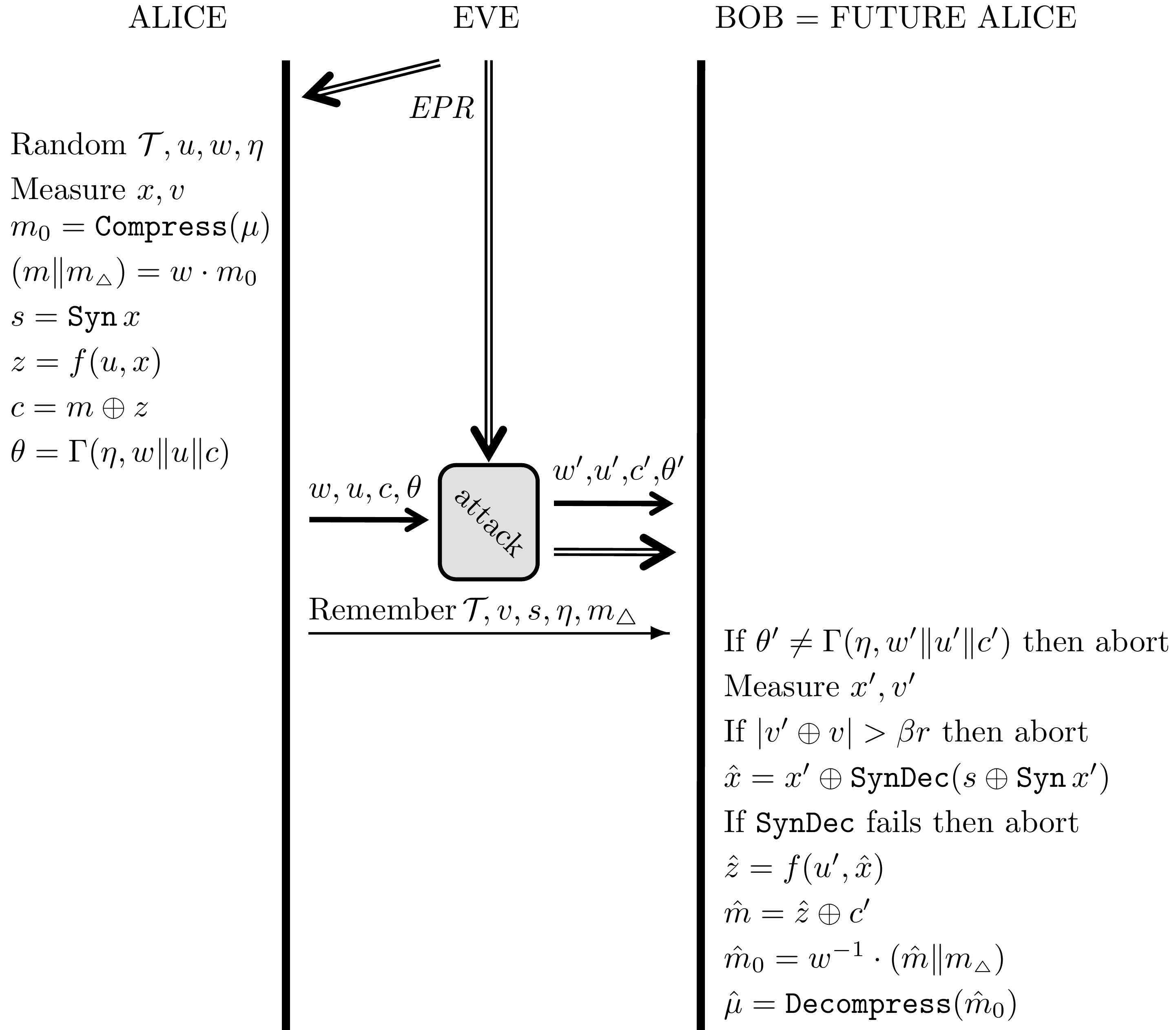}
\caption{\it 
The EPR version of {\sc Can'tTouchThis}. The double lines represent quantum states ($n+r$ qubits).
}
\label{fig:prot}
\end{center}
\end{figure}

\clearpage


\section{Security analysis}
\label{sec:proof}

\subsection{EPR version of the protocol}
\label{sec:EPR}

We present the EPR-pair based version of {\sc Can'tTouchThis} (Fig.\ref{fig:prot}). 
Only the differences with respect to Section~\ref{sec:protocolsteps} are listed.

Eve creates $n+r$ EPR-pairs.
Of each EPR pair she gives one qubit to Alice and keeps one for herself.
In step~3 of the protocol,
instead of preparing a state, Alice now measures her $j$'th qubit 
in the Hadamard basis if $j\in\cT$, and in the standard basis otherwise.
The random string $\xi$ is now the result of Alice's measurement.

In step~6 and later, we refer to Alice as `Bob'.

\subsection{Main result}
\label{sec:mainresult}

\begin{theorem}[Main theorem]
\label{th:main}
Consider the EPR version of {\sc Can'tTouchThis} as described in Section~\ref{sec:EPR}, 
with parameter values as in Section~\ref{sec:protocolsteps}. 
Let the distribution of the message $\mu$ and Alice's knowledge about this distribution be such
such that the length $\ell$ in (\ref{extractable}) is positive.
Let 
$\qd_c\isdef e^{-2(\qb-\qb_0)^2 r} + 2e^{-2(\qb+\nu-\qb_0)^2 n}$
and
$\qd\isdef	\exp\left[-2\nu^2 r \frac{nr}{(n+r)(r+1)}\right]$.\\
The protocol satisfies 
$\qd_c$-Correctness as defined in (\ref{defcorrect})
and is 
$(2\qe_{\rm mac}+2\qd+4\qe_0+2\qe_{\rm qp})$-Secure
as defined in Section~\ref{sec:securitydef}.
\end{theorem}

The security of the EPR version implies security of the actual protocol.

\underline{Proof of Theorem~\ref{th:main}}.\\
{\bf Correctness}.\\
If Eve is passive then the classical data $w,u,c,\qy$ remains unmodified.
The tag $\qy$ is automatically valid. The test in step~6 of the protocol
is passed with certainty. 
The qubits are subject to independent noise with bit error probability~$\qb_0$.
We will upper-bound the probability 
$\pr[\neg(\qO=1 \;\wedge\; \hat M=M)]$.
We write BER(data) for the bit error rate in the data positions, and similarly
BER(trap) for the trap positions.
\bea
	\pr[\neg(\qO=1 \;\wedge\; \hat M=M)] &=& \pr[\qO=0 \lor \hat M\neq M]
	\nn\\ &=&
	\pr[\qO=0]+\pr[\qO=1\land \hat M\neq M]
	\nn\\ &<&
	\pr[\qO=0]+\pr[\hat M\neq M]
	\nn\\ &\leq&
	\pr[\qO=0]+\pr[{\rm BER(data)}>\qb+\nu].
\label{corrderiv1}
\eea
In the last line we used that the event $\hat M\neq M$ can occur only if the bit
error rate in the non-trap qubits is too large for the ECC.

The event $\qO=0$ occurs either if the bit error rate in the trap positions exceeds $\qb$
or if the trap positions are OK but the SynDec fails in Step~8 of the protocol.
The {\tt SynDec} fails if the number of bit errors in $x'$ exceeds $n(\qb+\nu)$
{\em and} $s+{\tt Syn}\,x'$ happens to be a value that cannot be decoded.
\bea
	\pr[\qO=0] &=& \pr[{\rm BER(trap)}>\qb] + \pr[{\rm BER(trap)}\leq \qb \;\land\; \mbox{{\tt SynDec} fails}]
	\nn\\ &<&
	\pr[{\rm BER(trap)}>\qb] + \pr[\mbox{{\tt SynDec} fails}]
	\nn\\ &<&
	\pr[{\rm BER(trap)}>\qb] + \pr[{\rm BER(data)}>\qb+\nu].
\label{corrderiv2}
\eea
Substitution of (\ref{corrderiv2}) into (\ref{corrderiv1}) gives
\bea
	\pr[\neg(\qO=1 \;\wedge\; \hat M=M)] 
	& < &
	\pr[{\rm BER(trap)}>\qb] + 2\pr[{\rm BER(data)}>\qb+\nu]
	\nn\\ &\stackrel{{\rm Lemma}\;\ref{lemma:Hoeffding}}{<}&
	e^{-2(\qb-\qb_0)^2 r} + 2e^{-2(\qb+\nu-\qb_0)^2 n}.
\eea

{\bf Security}.\\
We have to prove that 
$\| \qr^{MTT'\rm E}_{[\qo=1]}-\qr^M\otimes\qr^{TT'\rm E}_{[\qo=1]} \|_1 \leq 2\qe_{\rm mac}+2\qd+4\qe_0+2\qe_{\rm qp}$.
First we note that $\pr[\qO=1 |T'\neq T]\leq \qe_{\rm mac}$.
This allows us to write
\be
	\| \qr^{MTT'\rm E}_{[\qo=1]}-\qr^M\otimes\qr^{TT'\rm E}_{[\qo=1]} \|_1
	\leq 2\qe_{\rm mac}+\| \qr^{MTT'\rm E}_{[\qo=1,t'=t]}-\qr^M\otimes\qr^{TT'\rm E}_{[\qo=1,t'=t]} \|_1 .
\ee
Next we write 
\be
	\| \qr^{MTT'\rm E}_{[\qo=1,t'=t]}-\qr^M\otimes\qr^{TT'\rm E}_{[\qo=1,t'=t]} \|_1
	\leq \| \qr^{M\hat MTT'\rm E}_{[\qo=1,t'=t]}-\qr^{M\hat M}_{[\qo=1,t'=t]}\otimes\qr^{ TT'\rm E}_{[\qo=1,t'=t]} \|_1.
\ee
(Taking the $\hat M$-trace cannot increase the norm.)
We apply Lemma~\ref{lemma:sampling} to the noise $v'\oplus v\in\bits^r$ in the trap positions
and the noise $x'\oplus x\in\bits^n$ in the data positions. 
Lemma~\ref{lemma:sampling} gives
\be
	\pr\Big[|v'\oplus v|\leq r\qb \;\;\wedge\;\;  |x'\oplus x| > n(\qb+\nu)\Big] \leq \qd.
\ee
It follows that
$\pr[\qO=1 \land \hat M\neq M]\leq \qd$.
This allows us to write
\be
	\| \qr^{M\hat MTT'\rm E}_{[\qo=1,t'=t]}-\qr^{M\hat M}_{[\qo=1,t'=t]}\otimes\qr^{ TT'\rm E}_{[\qo=1,t'=t]} \|_1
	\leq2\qd +
	\| \qr^{M\hat MTT'\rm E}_{[\qo=1,t'=t,\hat m=m]}-\qr^{M\hat M}_{[\qo=1,t'=t,\hat m=m]}\otimes\qr^{ TT'\rm E}_{[\qo=1,t'=t,\hat m=m]} \|_1.
\ee
Next we note that, by Lemma~\ref{lemma:LHL}, the $M$ is $\qe_0$ removed from being uniform.
If $M$ were uniform, the transcript $T$ (containing the ciphertext $C=M\oplus Z$) would be independent of~$Z$.
Hence, the state $\qr^{M\hat MTT'\rm E}_{[\qo=1,t'=t,\hat m=m]}$ is $\qe_0$-close (in terms of trace norm) to a sub-normalised
state $\qs^{MTE}$ where $M$ is uniform. 
We can write 
\be
	\| \qr^{M\hat MTT'\rm E}_{[\qo=1,t'=t,\hat m=m]}-\qr^{M\hat M}_{[\qo=1,t'=t,\hat m=m]}\otimes\qr^{ TT'\rm E}_{[\qo=1,t'=t,\hat m=m]} \|_1
	\leq 4\qe_0 + \| \qs^{MTE}-\chi^M \otimes\qs^{TE} \|_1.
\ee
Next we write $T=(C,U)$ (where we neglect the tag $w$) and explicitly write out the dependence of the state on
$m$ and $c$ to obtain
\bea
	\qs^{MCUE} &=& \sum_{mc}\frac1{2^\ell}\pr[C=c|M=m]\ket{mc}\bra{mc}\otimes \qs^{UE}_{mc}
	\\ &=&
	\sum_{zc} \frac1{2^\ell} \pr[Z=z] \ket{z\oplus c,c}\bra{z\oplus c,c}\otimes\qs^{UE}_{zc}
\label{replacembyz}
	\\ &=&
	\sum_{zc} \frac1{2^\ell} \pr[Z=z] \ket{z\oplus c,c}\bra{z\oplus c,c}\otimes\qs^{UE}_{z}.
\label{independentofc}
\eea
To get (\ref{replacembyz}) we have used that conditioning on $m,c$ is the same as conditioning on $z,c$.
In (\ref{independentofc}) we have used that the uniformity of $M$ makes the $U$E subsystem (which through $X$ is correlated with $Z$) independent from $C$.
Furthermore
\be
	\chi^M\otimes \qs^{CUE} = \sum_{mc}\frac1{2^{2\ell}}\ket{mc}\bra{mc}\otimes\qs^{UE}
	=\sum_{zc}\frac1{2^{2\ell}}\ket{z\oplus c,c}\bra{z\oplus c,c}\otimes\qs^{UE}.
\label{factorisedMCUE}
\ee
From (\ref{independentofc}) and (\ref{factorisedMCUE}) we get
\be
	\| \qs^{MCUE}-\chi^M \otimes\qs^{CUE} \|_1 = \| \qs^{ZUE}-\chi^Z \otimes\qs^{UE} \|_1.
\ee
Finally we have to show that we satisfy the conditions for the
existence of a 
quantum-proof extractor (Lemma~\ref{lemma:existExt}), in order to obtain
$\| \qs^{ZU\rm E}-\chi^Z\otimes \qs^{U\rm E} \|_1\leq 2\qe_{\rm qp}$.
For this we need a lower bound on the min-entropy $\Hmin(X|{\rm E})_\qs$.
Lemma~\ref{lemma:entropicBB84} gives
\be
	\Hmin(X|{\rm E})_\qs \geq n-\Hmax(X'|{\rm B})_\qs.
\ee
The $\Hmax(X'|{\rm B})_\qs$ represents the amount of redundancy that Bob (who shares noisy EPR pairs with Alice) needs 
in order to reconstruct a measurement at Alice's side.
For the state $\qs$, which is associated with values $\qo=1$ and $\hat m=m$, 
this redundancy can be upper bounded \cite{TL2017} as $n-\qk$.
This yields 
$\Hmin(X|{\rm E})_\qs \geq n-(n-\qk)=\qk$.
By the setting of $n$ relative to $\ell$ (\ref{setECC})
the conditions for Lemma~\ref{lemma:existExt} are indeed met.
\hfill$\square$


\section{Setting the parameters}
\label{sec:parameters}

\subsection{Non-asymptotic example}
\label{sec:paramexample}

We consider large but finite~$n$.
By way of example
we set the four terms $4\qe_0$, $2\qe_{\rm mac}$, $2\qd$ and $2\qe_{\rm qp}$
in Theorem~\ref{th:main} each equal to $\qe/4$, where $\qe$ is some constant.
(This is one way of getting $\qe$-security.)
We furthermore demand that the correctness parameter $\qd_c$ equals $\qe$ too. 
Getting the desired values for $\qd$ and $\qd_c$ requires tuning the parameters
$r,n,\qb,\nu$ as a function of $\qb_0,\qe$.
This can be done e.g. with the following choice,
\begin{itemize}[leftmargin=4mm,itemsep=0mm,topsep=1mm]
\item
Set $\ell$ according to (\ref{extractable}) with $\qe_0=\qe/4$;
then set $\qk=\ell+4\log\fr8\qe-2$.
\item
Set 
\be
	r  >  (\fr12-\qb_0)^{-2}\cdot4\ln\fr8\qe.
\label{setr}
\ee
\item
Take $n>2r$ and large enough such that en error-correcting code can be chosen that has message size $\qk$ and
codeword size $n$ and which can correct bit error rate 
\be
	\qb_0+\sqrt{\frac1{2r}\ln\frac1{\qe-2\qe^2}} +\sqrt{\frac3{2r}\ln\frac8\qe}.
\label{setn}
\ee
\item
Set
\bea
	(\qb-\qb_0)^2 &=& \frac1{2r}\ln\frac1{\qe-2\qe^{n/r}}
\label{setbeta}
	\\ 
	\nu^2 &=& \frac{\ln(8/\qe)}{2r} (1+\frac1r)(1+\frac rn).
\label{setnu}
\eea
\end{itemize}
The inequality (\ref{setr}) ensures that $\qb+\nu$ does not exceed~$1/2$.
Note that the bit error rate (\ref{setn}) is tuned to be slightly larger than the value
$\qb+\nu$ that follows from (\ref{setbeta},\ref{setnu}), in order to get an expression that does not depend on~$n$.
Also note that it can be advantageous to increase $r$ so that the correctable bit error rate
$\qb+\nu$ stays close to $\qb_0$, which prevents $n$ from growing very large.

Below we verify that (\ref{setbeta},\ref{setnu}) indeed yields $\qd_c\leq\qe$ and $2\qd\leq \qe/4$. 
First, we have $\qd_c=e^{-2r(\qb-\qb_0)^2}+2e^{-2n(\qb-\qb_0+\nu)^2}< e^{-2r(\qb-\qb_0)^2}+2e^{-2n(\qb-\qb_0)^2}$.
Substitution of (\ref{setbeta}) yields
$\qd_c< \qe-2\qe^{n/r} +2(\qe-2\qe^{n/r})^{n/r} < \qe-2\qe^{n/r} +2\qe^{n/r} =\qe$.

Second, isolating $\nu$ from $\qd=	\exp\left[-2\nu^2 r \frac{nr}{(n+r)(r+1)}\right]$ gives
$\nu^2=\frac{(n+r)(r+1)}{2nr^2}\ln\fr1\qd$. Setting $\qd=\qe/8$ yields (\ref{setnu}).

The numerical constants in the above example are not tight.

\subsection{Asymptotics}
\label{sec:paramasymp}

We look at the asymptotics for $\sH_2(\mu)\to\infty$ (which implies $n\to\infty$) at constant~$\qe$.
We consider the example above, with
power-law scaling of the number of trap states,
$r\propto n^\qa$, where  $\qa\in[0,1]$.
The case $\qa=0$ means that $r$ is independent of~$n$.
The case $\qa=1$ means that $r$ is a constant fraction of~$n$.
Eqs.~(\ref{setbeta},\ref{setnu}) become
\bea
	(\qb-\qb_0)^2 &\to& \cO(\frac1{n^\qa}\ln\frac1\qe )
	\\
	\nu^2 &\to& \cO(\frac1{n^\qa}\ln\frac8\qe ).
\eea
For $\qa>0$, the codeword length $n$ asymptotically tends to
$\frac{\ell}{1-h(\qb_0)}$, where $h(\cdot)$ is the binary entropy function.
The size of the syndrome is $n-\qk$, which tends to
$\ell \frac{h(\qb_0)}{1-h(\qb_0)}$.

It can be advantageous to choose $\qa$ close to~1 so that the asymptotic regime
sets in quickly.

\underline{How much local storage Alice needs. Asymptotic Usefulness.}\\
The main data item that Alice needs to remember (store locally) is the syndrome $s\in\bits^{n-\qk}$.
Asymptotically the size of the syndrome is $\ell \frac{h(\qb_0)}{1-h(\qb_0)}$ bits, i.e. linear in the size of the message.
In case $\qa<1$ all the other stored data items are sublinear in $\ell$.

The second largest stored data item consists of the set $\cT$ of trap locations, which takes
$\log{n+r\choose r}$ bits of storage. For $\qa<1$ this scales as $\cO(\ell^\qa\ln \ell)$;
for $\qa=1$ it scales as $\cO(\ell)$.

The third one is the $r\propto \ell^\qa$ trap values.

The size of the MAC key is only logarithmic in $\ell$, namely $2\log\fr1{\qe_{\rm mac}}+2\log(\ell_0+d+\ell)$.

Finally there is $\mdel\in\bits^{\ell_0-\ell}$.

It is difficult to make general statements about the gap $\ell_0-\ell$.
However, with increasing message length
the average compressed size (without padding) approaches the Shannon entropy of the message,
meaning that $M_0$ becomes more uniform,
and the ratio $(\ell_0-\ell)/\ell_0$ becomes smaller. 
Furthermore, for sources that produce i.i.d.~symbols it is known \cite{RennerWolf2004} that
asymptotically the smooth R\'{e}nyi entropy approaches the Shannon entropy,
which has an advantageous effect on (\ref{extractable}).
For these reasons, we expect the overhead $\ell_0-\ell$ to be sub-linear in $\log|\cM|$.

Delegated storage has positive Usefulness (see Section~\ref{sec:securitydef})
only if Alice needs to remember fewer than roughly $\ell$ bits.
For $\qa<1$ this condition is asymptotically expressed as
\be
	\ell \frac{h(\qb_0)}{1-h(\qb_0)}< \ell,
\ee
i.e.\,the bit error rate $\qb_0$ must be small enough to satisfy $1-2h(\qb_0)>0$.
This threshold is (perhaps unsurprisingly) the same as the BB84 threshold for having a positive QKD rate.

\subsection{Recursive application of {\sc Can'tTouchThis}}
\label{sec:recursive}

At large $\qb_0$ Alice gains very little from our scheme.
The following method may improve that.
The storage of the syndrome $s$ itself can be Delegated using {\sc Can'tTouchThis}; 
then Alice has to remember only a fraction 
$\frac{h(\qb_0)}{1-h(\qb_0)}$ of the original size.
This principle can be applied recursively until (asymptotically) Alice's local storage needs are 
very small compared to $\ell_0$.
The number of stored qubits is then 
$\frac{\ell}{1-h(\qb_0)}[1+\frac{h(\qb_0)}{1-h(\qb_0)}+\{\frac{h(\qb_0)}{1-h(\qb_0)}\}^2+\cdots]$
$=\frac{\ell}{1-2h(\qb_0)}$; 
this formula is familiar: it is associated with the key rate of (efficient) QKD,
i.e.~the number of qubits needed to convey a $\ell$-bit message one-time-pad-encrypted
with a QKD key.

Due to the composability of our scheme we expect that after $i$ iterations
the security has degraded from $\qe$ to~$i\qe$.
A proof is left for future work.

\section{Why the message needs to be randomised}
\label{sec:imposs}

The security proof (Section~\ref{sec:mainresult})  needs the assumption that either
the {\em fully randomised} or {\em weakly randomised} scenario holds (see Section~\ref{sec:securitydef}).
That leaves the question: Is it {\em impossible in general} to achieve Delegated Storage 
in the {\em non-randomised} scenario,
or is it just a quirk of our scheme and/or our proof method?

\subsection{The non-randomised scenario}
\label{sec:attackspecifics}

{\bf We consider a scenario where the distribution $P$ of the message is not known to Alice and is controlled by Eve.}
This is essentially the setting of Indistinguishability under Chosen Plaintext Attacks. 

Alice's lack of knowledge about $P$ prevents her from applying the prefix-code preprocessing trick, i.e.\,{\em any randomisation 
that she performs comes at the cost of having to remember keys, and hence becomes part
of the encryption procedure.}

By explicitly constructing an attack (we call it {\sc Support})
we demonstrate that, in this scenario, Delegated Storage with short keys is impossible.

We consider a general protocol, not restricted to the one proposed in Section~\ref{sec:protocolsteps}.
We use the following notation.
The plaintext is a random variable $M\in\cM$ with distribution $P$. 
We write $p_m\isdef\pr[M=m]$.
We denote the highest-probability plaintext as $m_*$, with probability $p_*$. 
Let $\rho(m,k)$ denote the quantum encryption of message $m$ using key $k$. 
The $\rho(m,k)$ represents {\em everything} (quantum and classical) that Alice stores on the server, 
while $k$ is everything that Alice stores privately. 
The stored state does not have to be pure.
The event that Alice does not notice disturbance is called {\tt acc} (`accept'),
and if she notices disturbance {\tt rej} (`reject').
We consider only `correct' schemes, i.e. decryption succeeds with certainty when there is no attack.

\subsection{The {\sc Support} Attack}
\label{attack}

Alice receives $m$ from the distribution $P$ and draws a key $k$ uniformly from~$\cK$.
She creates the encryption $\qr(m,k)$.
Eve's task is to determine from the encryption whether $m=m_*$ without causing a {\tt rej}.
We set some notation:
\begin{itemize}[leftmargin=4mm,itemsep=0mm,topsep=1mm]
\item
For $y\in\cM$
we write $\qr(\neg y,k)\isdef\frac1{1-p_y}\sum_{m\neq y}p_m \qr(m,k)$
and $\qr(\cM,k)\isdef\sum_{m\in\cM}p_m\qr(m,k)$.
\item 
For $m\in\cM$, $k\in\cK$
let $\Pi_{m,k}$ be the projector onto $\operatorname{support}\left(\rho(m,k)\right)$.
\item 
For $m\in\cM$
let $\operatorname{\Pi}_{m,\cK}$ be the projector onto 
$\operatorname{support}\left(\sum_{k\in\mathcal{K}}\operatorname{\Pi}_{m,k}\right)$.
\item 
For $k\in\cK$
let $\operatorname{\Pi}_{\mathcal{M},k}$ be the projector onto 
$\operatorname{support}\left( \sum_{m\in\cM} \operatorname{\Pi}_{m,k} \right)$.
For the correctness of the decryption the orthogonality property
$\operatorname{\Pi}_{\mathcal{M},k}=\sum_{m\in\cM}\operatorname{\Pi}_{m,k}$ must hold.

\item 
Let $\operatorname{I}_{\mathcal{M},\mathcal{K}}$ be the (identity) projector onto 
$\operatorname{support} \left( \sum_{m\in\cM} \operatorname{\Pi}_{m,\mathcal{K}} \right)$.
\end{itemize}

\begin{definition}
The attack {\sc Support} proceeds as follows:
Eve applies, on the quantum state she receives from Alice, 
the projective measurement 
$\{\operatorname{\Pi}_{m_*,\cK},\;\operatorname{I}_{\cM,\cK}-\operatorname{\Pi}_{m_*,\cK}\}$. 
If she obtains $\operatorname{\Pi}_{m_*,\cK}$, she guesses $m_*$; otherwise, she guesses $\neg m_*$.
\end{definition}

The  projective measurement needs to keep all the ciphertexts in superposition;
it is potentially difficult to implement as it needs a (quantum computational) operation with 
entanglement of many qubits.

\subsection{{\sc Support} breaks the security in the non-randomised scenario}

We denote by $\operatorname{WIN}$ the event that 
Eve guesses correctly whether $m=m_*$.
 
\begin{lemma}
For $m= m_*$ the {\sc Support} attack causes {\rm WIN} and {\tt acc}.
\end{lemma}
\underline{\it Proof:}
In the case that Alice encrypts $m= m_*$, no matter the actual key $k$ used, the {\sc Support} measurement 
$\Pi_{m_*,\cK}$
leaves the state unchanged and correctly yields the measurement result $m_*$.
\hfill$\square$

\begin{lemma}
\label{lemma:globalacc}
For the overall {\tt acc} probability we have
	$\pr[{\tt acc}]\geq p_*$\,.
\end{lemma}
\underline{\it Proof:}
$\pr[{\tt acc}]=\pr[M=m_*] \pr[{\tt acc}|M=m_*]
+\pr[M\neq m_*] \pr[{\tt acc}|M\neq m_*]
  \ge 
p_* \pr[{\tt acc}|M=m_*]
= p_*$.
\hfill$\square$

We introduce the following notation.
Let $P$ be a distribution on $\cM$, and let $\pi$ be a permutation on $\cM$.
The permuted distribution is denoted as $\pi(P)$.

\begin{lemma}
\label{lemmasmall}
Consider the scenario described in Section~\ref{sec:attackspecifics}.
Let $P$ be a distribution on $\cM$.
There exists a permutation $\pi$ on $\cM$ such that
the {\sc Support} attack has 
$
\pr_{M\sim \pi(P)}[ {\rm WIN} | M\neq m_*] \ge 1-\frac{| \cK |}{| \cM |}
$.
\end{lemma}

\underline{\it Proof:}
We use 
$\max_{\pi}\pr_{M\sim \pi(P)}[ {\rm WIN} | M\neq m_*] \geq $
$\EE_{\pi} \pr_{M\sim \pi(P)}[ {\rm WIN} | M\neq m_*]$.
In the derivation below the effect of $\EE_{\pi}$ is a uniform choice of $m_*$.
For all $k\in\cK$ we have
\bea
	&& \EE_\pi \pr_{M\sim \pi(P)}[\mbox{WIN}|M\neq m_*,k] = 
	1-\frac1{|\cM|}\sum_{m_*\in\cM} \tr   \Pi_{m_*,\cK} \cdot \rho(\neg m_*,k)
	\nn\\ && \quad\quad\quad\quad\quad = 
	1-\frac1{|\cM|}\sum_{m_*\in\cM} \tr   \Pi_{m_*,\cK} \cdot \frac1{1-p_*}\sum_{m\neq m_*}p_*\qr(m,k)
	\nn\\ && \quad\quad\quad\quad\quad =
	1-\frac1{|\cM|}\sum_{m_*\in\cM} \frac1{1-p_*}[\tr \Pi_{m_*,\cK}\cdot \qr(\cM,k)-p_* ]
\label{XP1}
	\\ && \quad\quad\quad\quad\quad \geq
	1-\frac1{|\cM|}\sum_{m_*\in\cM} \tr   \Pi_{m_*,\cK} \cdot \rho(\cM,k)
\label{XP2}
	\\ && \quad\quad\quad\quad\quad \geq 
	1-\frac1{|\cM|}\sum_{m_*\in\cM} \sum_{k'}\tr   \Pi_{m_*,k'} \cdot \rho(\cM,k)
\label{XP3}
	\\ && \quad\quad\quad\quad\quad =
	1- \sum_{k'}\frac1{|\cM|}\sum_{m_*\in\cM} \tr   \Pi_{m_*,k'} \cdot \rho(\cM,k)
	\\ && \quad\quad\quad\quad\quad \stackrel{\rm correctness}{=}
	1-\sum_{k'}\frac1{|\cM|}\tr \Pi_{\cM,k'}\cdot \rho(\cM,k)
	\\ && \quad\quad\quad\quad\quad \geq
	1-\frac{|\cK|}{|\cM|}.
\eea
The equality (\ref{XP1}) follows from the definitions of $\qr(\cM,k)$, $\qr(\neg m_*,k)$
and the fact that $\tr [\Pi_{m_*,\cK}$ $\qr(m_*,k)]$ $=1$.
The inequality (\ref{XP2}) follows from 
$p_*\geq p_* \tr   \Pi_{m_*,\cK} \cdot \rho(\cM,k)$.
We get (\ref{XP3}) from 
$\sum_{k’} \Pi_{m_*,k’}$
$\geq  \Pi_{m_*, \mathcal{K}}$.
\hfill$\square$

\begin{lemma}
\label{lemmasmall2}
Consider the scenario described in Section~\ref{sec:attackspecifics}.
Let $P$ be a distribution on $\cM$.
There exists a permutation $\pi$ on $\cM$
such that
\be
	\pr_{M\sim\pi(P)}[{\tt acc}|M\neq m_*] \le 
	\pr_{M\sim\pi(P)}[{\rm WIN}\land {\tt acc}|M\neq m_*] + \frac{|\cK|}{|\cM|}.
\ee
\end{lemma}

\underline{\it Proof:}
We have 
$\pr[{\rm WIN}\land {\tt acc}|M\neq m_*]
= 
\pr[{\rm WIN}|M\neq m_*]-\pr[{\rm WIN}\land \neg {\tt acc}|M\neq m_*]$.
Applying Lemma~\ref{lemmasmall} gives
\begin{align}
\exists_\pi\quad \pr_{M\sim\pi(P)}[{\rm WIN}\land {\tt acc}|M\neq m_*]
& \geq
1- \frac{|\cK|}{|\cM|}-\pr_{M\sim\pi(P)}[\operatorname{WIN\land \neg {\tt acc}}|M\neq m_*] 
\\ &\ge
1- \frac{|\cK|}{|\cM|}-\pr_{M\sim\pi(P)}[\neg{\tt acc}|M\neq m_*]
\\ &=
\pr_{M\sim\pi(P)}[{\tt acc}|M\neq m_*]- \frac{|\cK|}{|\cM|}\,.
\end{align}\hfill$\square$

\begin{proposition}
\label{prop:attack}
Consider the scenario described in Section~\ref{sec:attackspecifics}.
Let $P$ be a distribution on $\cM$.
There exists a permutation $\pi$ on $\cM$
such that
the {\sc Support} attack has the following advantage in guessing the bit $[m=m_*]$ correctly
while staying unnoticed by Alice.
\be
	\pr_{M\sim\pi(P)}[{\rm WIN}|{\tt acc}]-p_* \;\ge\; p_*(1-p_*)(1-\frac{|\mathcal{K}|}{|\mathcal{M}|}).
\label{advantage}
\ee
\end{proposition}

\underline{\it Proof:}
We have
\bea
\pr[{\rm WIN}|{\tt acc}] 
&=&
\pr[{\rm WIN}\land {\tt acc}]/\pr[{\tt acc}]
\\&= &
\frac{p_*\cdot \pr[{\rm WIN}\land {\tt acc}|M=m_*]+(1-p_*)\cdot 
\pr\left[{\rm WIN}\land {\tt acc} |M\neq m_*\right]}{p_*\cdot \pr[{\tt acc}|M=m_*]
+(1-p_*)\cdot \pr[{\tt acc}|M\neq m_*]}
\\ &=&
\frac{p_*+(1-p_*)\cdot \pr\left[{\rm WIN}\land {\tt acc} |M\neq m_*\right]}
{p_* +(1-p_*)\cdot \pr[{\tt acc}|M\neq m_*]}
\eea
Note that $\pr[{\tt acc}]\neq 0$ by Lemma~\ref{lemma:globalacc}, allowing the division by $\pr[{\tt acc}]$ in the first line.
Applying Lemma~\ref{lemmasmall2} yields
\bea
\exists_{\pi}\quad \pr_{M\sim\pi(P)}[{\rm WIN}|{\tt acc}] &\ge &
\frac{p_*+(1-p_*)\pr_{M\sim\pi(P)}\left[ {\rm WIN}\land {\tt acc} | M\neq m_*\right]}
{p_*+(1-p_*)\{ \pr_{M\sim\pi(P)}[{\rm WIN}\land {\tt acc}|M\neq m_*] + \frac{|\cK|}{|\cM|} \} }
\\&\ge & 
\frac{p_*}{p_*+ (1-p_*)\frac{|\mathcal{K}|}{|\mathcal{M}|}}
=
\frac{p_*}{1-(1-p_*)(1-\frac{|\mathcal{K}|}{|\mathcal{M}|})}
\\ &\geq &
p_* + p_*(1-p_*)(1-\frac{|\mathcal{K}|}{|\mathcal{M}|}).
\eea
\hfill$\square$

For the non-randomised scenario,
Proposition~\ref{prop:attack} and Lemma~\ref{lemma:globalacc} prove 
that the security property (\ref{secstatement}) cannot be achieved with a short key.
Consider a distribution $P$ such that $p_*\gg 1/|\cM|$.
In order to make Eve's advantage (\ref{advantage}) negligible
it is necessary to make the key length almost equal to the message length;
this negates all the advantages of delegated storage.

\begin{theorem}
\label{th:nonexistDS}
For all $Y>0$ there exists $\qe >0$ such that there exists no
$Y$-useful, $\qe$-secure Delegated Storage protocol in the non-randomised scenario.
\end{theorem}
\underline{\it Proof}.
By Proposition~\ref{prop:attack} the attacker's advantage is lower bounded by
$p_*(1-p_*)Y$, with $Y$ constant as a function of the message size.
In the non-randomised scenario, the attacker controls $p_*$,
so Alice has no way to reduce the attacker's advantage below $p_*(1-p_*)Y$.
(E.g. increasing the message length, which typically improves security, does not help here.)
The fixed advantage does not allow $\qe$ to be decreased indefinitely. 
\hfill $\square$

\section{Discussion}
\label{sec:discussion}

Various alternative constructions are of course possible.
Better parameter values may be obtained. 
For instance, as the quantum-proof strong extractor one could use a Trevisan extractor
\cite{Trevisan2001,DPVR2012} would yield a seed of length
$\cO(\log \ell \log^2(\fr n\qe))$ instead of length~$n$.

If confidentiality is required in case of a {\tt reject}, Alice can classically encrypt the message,
before or after randomisation, or as part of the randomisation.
The confidentiality will not be information-theoretic since the encryption key has to be short.

A different way to improve the {\tt reject}-case
confidentiality is to do secret sharing of the message between two or more servers. 
Then the plaintext is compromised only if 
(i) all the retrievals are {\tt reject}s; and
(ii) all servers collude.

Encryption can be used for a different purpose as well.
Consider the scenario discussed in Section~\ref{sec:imposs}, i.e.~Alice is unable to randomise the message $\mu$ `for free'.
Let Alice use a classical cipher $F$ to create ciphertext $\nu=F_k(\mu)$, where $k$ is a short key.
This gives Alice a benefit: from Eve's point of view,  
the $\nu$ {\em temporarily} looks random (until Eve breaks $F$), enabling Alice to apply {\sc Can'tTouchThis}
with $\nu$ as the message to be stored.
Thus, the tamper evidence in this scenario
is computational instead of information-theoretic;
still a feat that cannot be accomplished classically.
Furthermore, Eve works under a time limit. 
She is forced break $F$ before Alice retrieves the data.

Secret sharing over multiple servers can achieve this message-randomisation purpose too, 
and the servers only need to be prohibited from colluding {\em during} the protocol, 
because the security granted by tamper evidence is unaffected by the servers sharing information {\em after} the verification phase.

An interesting aspect of our results is that unconditional tamper evidence for the randomised version of delegated storage 
does not imply unconditional tamper evidence for its non-randomised version. 
In fact, Theorem~\ref{sec:mainresult} and Theorem~\ref{th:nonexistDS} respectively prove that, given a requirement of non-zero usefulness 
for at least some large-enough messages, the former is possible while the latter is not.
In contrast, the security of non-randomised Oblivious Transfer~\cite{beaver1995precomputing} 
has been shown, under many sequential composability scenarios~\cite{crepeau2006information,wehner2008composable,fehr2009composing}, 
to be reducible to the security of Randomised Oblivious Transfer.
Similarly, Quantum Key Distribution has been shown~\cite{ben2005universal} secure 
in the universally composable sense: 
In order to communicate non-random messages, the random keys that QKD distributes can be securely correlated with subsequent information.

There is some resemblance between Delegated Storage (DS) and Unclonable Encryption (UE) \cite{uncl,LS2021}, 
in that a special security guarantee is given
only in the `accept' case, and that the keys can be revealed after an `accepted' completion of the protocol.
However, there is one significant difference:
In DS Alice's keys have to be small, whereas UE needs the key to be as long as the message.
Thus UE could be seen as DS with bad Usefulness.
UE with computational confidentiality instead of unconditional confidentiality would come closer to DS.

\vskip2mm

An interesting extension of our construction would be to apply {\sc Can'tTouchThis} to {\em quantum information}.
The trap qubits would work in a similar way; the mask $z$ would become a key for
Quantum-One-Time-Pad encryption of Alice's quantum information.

\vskip4mm

{\bf\Large Acknowledgements}\\

We thank Daan Leermakers, Dave Touchette, and Thomas Vidick for useful discussions and references.
Part of this work was funded by
Fonds de Recherche du Qu\'{e}bec -- Nature et Technologies (FRQNT) and the Swiss National Science Foundation (SNF).

\bibliographystyle{unsrt}
\bibliography{delegated_arxiv2}


\end{document}